\documentclass{sig-alternate}
\usepackage{wrapfig}

\begin{document}

\title{Feasibility of Using Automatic Speech Recognition with Voices of Deaf and Hard-of-Hearing Individuals}

\CopyrightYear{2017} 
\setcopyright{rightsretained} 
\conferenceinfo{ASSETS '17}{October 29-November 1, 2017, Baltimore, MD, USA}\isbn{978-1-4503-4926-0/17/10}
\doi{https://doi.org/10.1145/3132525.3134819}

\clubpenalty=10000 
\widowpenalty = 10000

\author{
\alignauthor
		Abraham Glasser, Kesavan Kushalnagar, and Raja Kushalnagar\\
       \affaddr{Rochester Institute of Technology}\\
       \affaddr{Rochester, NY 14623}\\
       \email{\{atg2036, krk4565\}@rit.edu, raja.kushalnagar@gallaudet.edu}
}

\maketitle





\abstract{
Many personal devices have transitioned from visual-controlled interfaces to speech-controlled interfaces to reduce costs and interactive friction, supported by the rapid growth in capabilities of speech-controlled interfaces, e.g., Amazon Echo or Apple's Siri. A consequence is that people who are deaf or hard of hearing (DHH) may be unable to use these speech-controlled devices. We show that deaf speech has a high error rate compared to hearing speech, in commercial speech-controlled interfaces. Deaf speech had approximately a 78\% word error rate (WER) compared to a hearing speech 18\% WER. Our findings show that current speech-controlled interfaces are not usable by DHH people. 
Based on our findings, significant advances in speech recognition software or alternative approaches will be needed for deaf use of speech-controlled interfaces. We show that current speech-controlled interfaces are not usable by DHH people.
}




\section{Related Work}
Prior research has investigated how the lack of a feedback loop for deaf people who cannot hear their own speaking results in poor speech quality due vowel errors, intonation errors, and length errors~\cite{thirumalai_gayathri_2004,hearingimpairedspeech}. ASR software is generally trained with speech samples from hearing people, which results in very poor recognition of deaf speech. Even when used with limited possibilities, e.g., single digits, ASR for deaf speakers with poor speech intelligibility yielded a 13\% Word Error Rate (WER)~\cite{jeyalakshmi2010deaf}, compared to nearly no errors for hearing speech \cite{doddington1989phonetically}.

\begin{figure}
	\centering
	\includegraphics[width=0.5\textwidth]{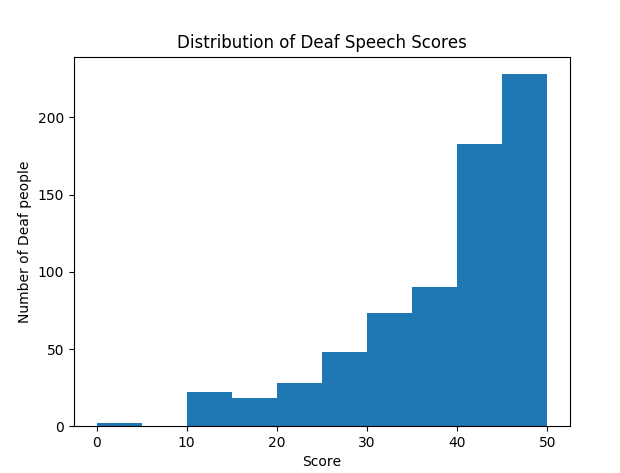}
    \caption{Distribution of Clarke Sentence Scores.}
    \label{fig:Distribution of Clarke Sentence Scores}
\end{figure}
\vspace{-1em}

\section{Methodology}

\subsection{Deaf Speech Dataset}


We sampled from a subset of a large speech dataset of 650 deaf and hard of hearing (DHH) individuals at the National Technical Institute for the Deaf at Rochester Institute of Technology, which has an enrollment of around 1100 deaf and hard of hearing students. The dataset consisted of samples taken from DHH individuals who took the Clarke Sentences intelligibility test ~\cite{doi:10.1044/jshr.3103.307}. The test has 60 sentence lists, each with 10 sentences of 10 syllables. The number of words varies across the sentences and lists. Each audio file has one speaker speaking one list from the Clarke sentence list. In each audio file, the speaker says the sentence number and then proceeds to say that sentence, and repeats until all the ten sentences are spoken. The audio files were recorded by one individual, then the samples were sent to a speech pathologist. The speech pathologist assigned an intelligibility score from 0 to 50. The score is computed by looking for 50 target words within the sample for credit. A score of 50 would indicate that the deaf person is generally intelligible to the speech pathologist, while a score of 30 means difficult to understand, and a score of 0 means completely unintelligible. About half of all deaf individuals had scores of under 40, which is usually unintelligible to people not used to deaf speech, as shown in Figure~\ref{fig:Distribution of Clarke Sentence Scores}. 

\subsection{ASR and WER Analysis}
We used the Microsoft Translator Speech API to create transcriptions for each audio file. This API is used by businesses for transcriptions. As a commercial level software, it matches other similar transcription software \cite{eckel_2016,venturebeat_2017,saon_2017}.
We also used the National Institute of Standards and Technology Speech Recognition Scoring Toolkit (SCTK) Version 2.4.0.4 for the Word Error Rate calculations~\cite{nist_2017}. 




\begin{figure}
    \includegraphics[width=0.5\textwidth]{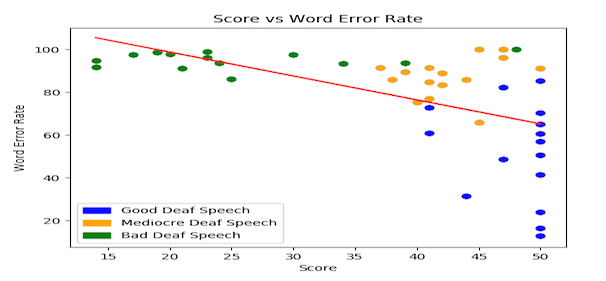}
    \vspace{-1em}
    \label{fig:Deaf Speech}
    \caption{Intelligibility rating vs Word Error Rate}
\end{figure}

\section{Analysis Results}


\subsection{WER for Hearing Speakers}

We calculated the ASR transcription and WER analysis results for five hearing subjects who read various lists from the same Clarke Sentences database, The speech samples were recorded with a cell phone in a noisy environment with background noise in a lab with many people speaking and computers, which is similar to common use-case scenarios in using phone ASR services.
The average WER was 18\%. While the speech recognition was not close to perfect for hearing people, it was still passable. These numbers are expected from the current state of the art technology.  
The majority of voice command interfaces are currently found in cell phones and home assistants, which are often used in noisy environments.
If we had repeated these recordings in the same setting as the deaf speech samples, we would expect an even lower WER.

\subsection{WER for Deaf Speakers}
We ran a sample of the deaf speech database through the Microsoft Translator Speech API. We used 45 total samples that were chosen by a naive listener who determined 15 good samples (40+), 16 mediocre samples (30-40), and 14 poor files (10-30). 
The error rates were extremely high over all samples at 77\%, including 53\% WER for the good samples. 

\begin{figure}
    \includegraphics[width=0.5\textwidth]{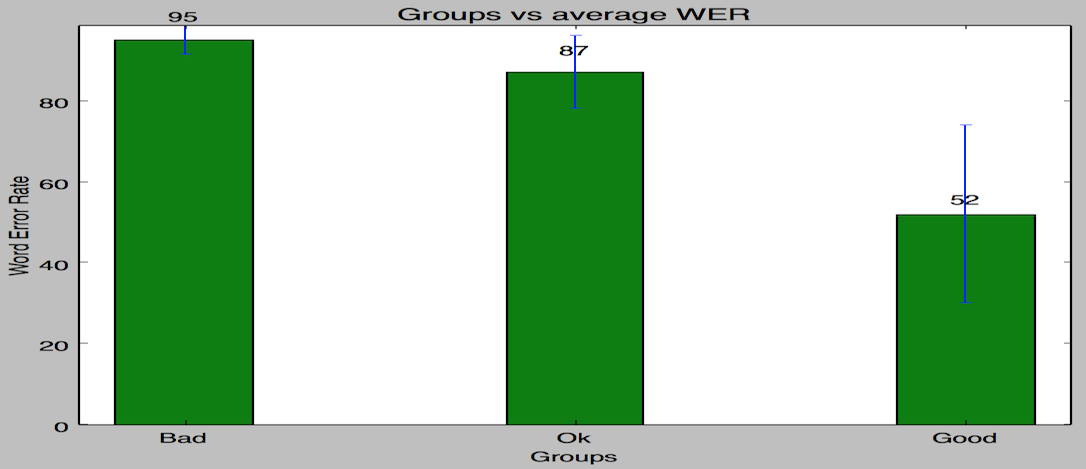}
    \label{fig:Statistics}
    \vspace{-1em}
    \caption{Group vs average WER}
\end{figure}

The average WER and standard deviation was calculated for each group, as shown in Figure~\ref{fig:Statistics}. The average WER for the good speech group was significantly less than either the mediocre group or bad speech groups; a t-test comparison between the good and mediocre groups yielded p < 0.01.

\section{Conclusions}
The current WER of Microsoft Translator was too high for comfortable use. They were significantly poorer in performance compared to hearing people under similar conditions. There are a number of factors that have an impact on the accuracy of automatic speech recognition systems with deaf speech.
The results also show much greater variance among deaf speakers, compared with hearing speakers. In order for ASR systems to recognize deaf speech as well as it does hearing speech requires a huge database of deaf speakers. While conceptually simple, it is still challenging. The deaf population is relatively small compared to the size of the hearing population, and have far more varied backgrounds.

ASR systems are trained using huge hearing speaker datasets. The results show that even deaf speakers with ``good'' speech have worse accuracy compared with the average hearing speaker. Although the Clarke Sentence test is useful for speech pathology evaluation, it is less useful for providing more feedback to DHH people about the usability of current ASR interfaces, since the top rating of 50 does not distinguish between DHH speakers with high ASR accuracy and those with less ASR accuracy. It would be helpful to develop an automated test that provides feedback to DHH people on their use of ASR services such as Siri or Alexa.  

\bibliographystyle{unsrt}
\bibliography{sigproc}

\end{document}